# Thermoelectric properties of vibrating molecule asymmetrically connected to the electrodes

Kamil Walczak

Department of Electrical & Computer Engineering, University of Virginia
351 McCormick Road, Charlottesville, Virginia 22904, USA

Here we present a theoretical analysis of inelastic effects on thermoelectric properties of molecular-scale junction in both linear and nonlinear response regimes. Considered device is composed of molecular quantum dot (with discrete energy levels) asymmetrically connected to metallic electrodes (treated within the wide-band approximation) via potential barriers, where molecular vibrations are modeled as dispersionless phonon excitations. Nonperturbative computational scheme, used in this work, is based on Green's function theory within the framework of polaron transformation (GFT-PT) which maps the many-body electron-phonon interaction problem into a one-body multi-channel single-electron scattering problem. It is shown that all the thermoelectric characteristics are dominated by quantum transport of virtual polarons due to a strong electron-phonon coupling.

Key words: thermoelectric effect, polaronic transport, electron-phonon interaction, molecular diode
PACS numbers: 85.65.+h, 73.23.Ad, 71.38.-k, 65.80.+n

## 1. Introductory remarks

In recent years, molecular devices are suggested to become components in future electronic circuits [1-4]. Such type of devices are composed of single molecules (or molecular layers) connected to two (or more) electrodes (reservoirs) and their transport properties are dictated by the chemical, physical and electronic structures of molecules. A molecule itself represents quantum dot with discrete energy levels, at least an order of magnitude smaller than semiconductor quantum dots. Usually contact with the electrodes is sufficiently weak and one can treat molecular quantum dot as electrically isolated from metallic electrodes via potential barriers. However, in contrast to rigid semiconductor dots, molecules involved into the conduction process can be thermally activated to vibrations at finite temperatures. The electrons passing through energetically accessible molecular states (conduction channels) may exchange a definite amount of energy with the nuclear degrees of freedom, resulting in an inelastic component to the current. Such molecular oscillations can have essential influence on the shape of transport characteristics especially in the case, when the residence time of a tunneling electron on a molecular bridge is of order of magnitude of the time involved in nuclear vibrations ($\sim ps$). So far, the influence of inelastic scattering processes on the current-voltage (I-V) characteristics [5-9] and shot noise [10-12] have been considered, while their influence on thermoelectric properties has not been addressed for molecular dots.

Anyway, a current between two reservoirs is not only related to the difference in the chemical potential, but it can be also related to the difference in temperature of considered reservoirs. Therefore, the thermoelectric effects can play a major role in molecular electronics, providing information on electronic and vibrational excitation spectrum of the molecule itself [13-16]. For example, it was suggested that a measurement of the thermoelectric voltage can provide new insights into molecular transport, giving information about the nature of conduction (electron or hole type) and allows one to estimate the Fermi energy relatively to the molecular energy levels [13]. In reality, the thermoelectric voltage was measured with a 20K temperature difference over a monolayer of Guanine molecules on a graphite substrate with the help of scanning tunnelling microscope (STM apparatus) [17]. However, measurements of thermoelectric effects still remain a certain challenge in molecular transport.

The main purpose of this work is to generalize existing nonperturbative method, based on Green's function theory within the framework of polaron transformation (GFT-PT) [5-9] to include both electrical and heat currents. Our treatment transforms the many-body electron-phonon interaction problem into a one-body multi-channel single-electron scattering problem (as presented graphically in Fig.1), while the many-electron nature of the process of molecular conduction is not considered (i.e. all the interactions between charge carriers are neglected). Anyway, it should be emphasized that the present approach is an exact method to deal with an interacting electron-phonon system having arbitrary coupling strength if the maximum number of phonons is taken up to infinity. In practice, we have to truncate this number to a finite value chosen to ensure computational convergence with desired accuracy. This choice is dictated by: the phonon energy, the strength of the electron-phonon interaction and the vibrational temperature of the molecular junction. In fact, the present method can be well-justified in the weak nonequilibrium thermal conditions or in the case, where the molecule is strongly connected to only one electrode (as expected in the STM-involved experiments). Moreover, discussed approach can provide a satisfactory description of the current spectrum consistent with actual experimental results [18-23].

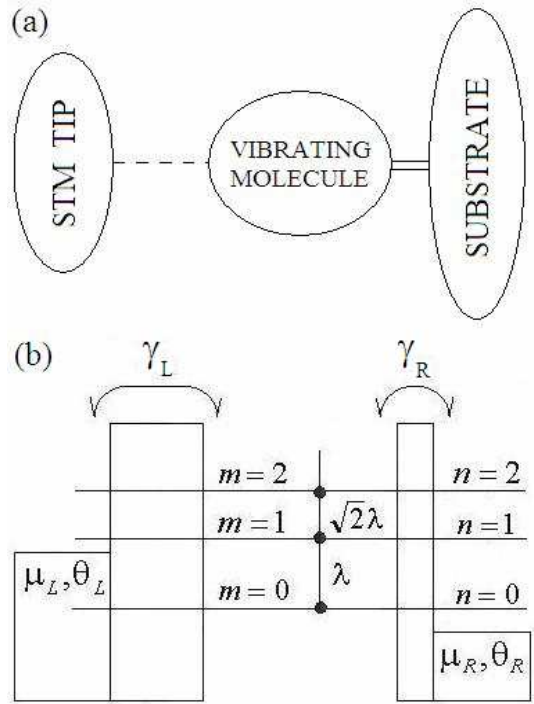

Figure 1: A schematic representation of the analyzed device composed of molecular quantum dot asymmetrically connected to two reservoirs via potential barriers (a) and the corresponding inelastic scattering problem via single energy level of the molecule (b).

## 2. Theoretical background

2.1 Description of the model and polaron transformation.

Let us hypothetically distinguish the three parts of the considered device, where the central molecular bridge is isolated from two reservoirs (left and right) via potential barriers, as shown in Fig.1. Molecular vibrations are modeled as dispersionless (optical) phonon excitations which can locally interact with conduction electrons. In this case, the full Hamiltonian of the system can be written as a sum:

$$H = \sum_{\alpha} H_{\alpha} + H_M + H_T . \qquad (1)$$

Here: $\alpha = L$ for the left electrode (STM tip) and $\alpha = R$ for the right one (substrate), respectively, in the case of two-terminal junction. Both metallic electrodes are treated as reservoirs for non-interacting electrons and described with the help of the following Hamiltonian:

$$H_L + H_R = \sum_{k \in \alpha} \varepsilon_k c_k^+ c_k . \qquad (2)$$

Here: $\varepsilon_k$ is the single particle energy of conduction electrons, while $c_k^+$ and $c_k$ denote the electron creation and annihilation operators with momentum $k$ in the $\alpha$-electrode. The third term in Eq.(1) represents molecular Hamiltonian:

$$H_M = \sum_i \left[ \varepsilon_i - \sum_j \lambda_j (a_j + a_j^+) \right] d_i^+ d_i + \sum_j \Omega_j a_j^+ a_j , \qquad (3)$$

Here: $\varepsilon_i$ is single energy level of the molecule, $\Omega_j$ is phonon energy in mode $j$, $\lambda_j$ is the strength of on-level electron-phonon interaction. Furthermore, $d_i^+$ and $d_i$ are electron creation and annihilation operators on level $i$, while $a_j^+$ and, $a_j$ are phonon creation and annihilation operators, respectively. The last term in Eq.(1) describes the molecule-electrode coupling:

$$H_T = \sum_{k \in \alpha; i} \left( \gamma_{k,i} c_k^+ d_i + h.c. \right), \qquad (4)$$

where the matrix elements $\gamma_{k,i}$ stands for the strength of the tunnel coupling between the dot and metallic electrodes.

The problem we are facing now is to solve a many-body problem with phonon emission and absorption when the electron tunnels through the dot, neglecting all the electron-electron interactions. Let us consider for transparency only one phonon mode (primary mode), since generalization to multi-phonon case can be obtained straightforwardly. The electron states inside the molecule are expanded onto the direct product states composed of single-electron states and $m$-phonon Fock states:

$$|i, m\rangle = d_i^+ \frac{(a^+)^m}{\sqrt{m!}} |0\rangle , \qquad (5)$$

where electron state $|i\rangle$ is accompanied by $m$ phonons ($|0\rangle$ denotes the vacuum state). Similarly the electron states in the electrodes can be expanded onto the states:

$$|k,m\rangle = c_k^+ \frac{(a^+)^m}{\sqrt{m!}}|0\rangle, \qquad (6)$$

where the state $|k\rangle$ with momentum $k$ is accompanied by $m$ phonons. Using the mentioned polaron transformation, the non-interacting single-mode electrodes (Eq.(2)) are mapped to a multi-channel model:

$$\tilde{H}_L + \tilde{H}_R = \sum_{k\in\alpha;m}(\varepsilon_k + m\Omega)|k,m\rangle\langle k,m|. \qquad (7)$$

Since the molecule is strongly connected to only one electrode (substrate), let us assume that the molecule is in the state of thermal equilibrium with the mentioned reservoir. Therefore the vibrational population (i.e. accessibility of particular conduction channels) is approximately determined by a weight factor:

$$P_m = [1 - \exp(-\beta_R \Omega)]\exp(-m\beta_R \Omega). \qquad (8)$$

Here Boltzmann distribution function is used to indicate the statistical probability of the phonon number state $|m\rangle$ at finite temperature $\theta_R$ of the $R$-reservoir, $\beta_R^{-1} = k_B \theta_R$ and $k_B$ is Boltzmann constant. Here we neglect the influence of nonequilibrium effects on molecular vibrations that could be crucial in the case of symmetric molecule-electrode connection, where the correct population of vibrational levels should be accounted self-consistently. It is clear that the derivatives of $P_m$ with respect to temperature are not included, since $\beta_R$ is kept fixed. Neglecting also the dissipative processes, the electron energies are constrained by the following energy conservation law:

$$\varepsilon_{in} + m\Omega = \varepsilon_{out} + n\Omega, \qquad (9)$$

where $\varepsilon_{in}$ is the energy of the incoming electron with the initial amount of phonons $m$, while $\varepsilon_{out}$ is the energy of outgoing electron with the final amount of phonons $n$, respectively. In the new representation (Eq.(5)), molecular Hamiltonian (Eq.(3)) can be rewritten in the form:

$$\tilde{H}_M = \sum_{i,m}(\varepsilon_i + m\Omega)|i,m\rangle\langle i,m| - \sum_{i,m}\lambda\sqrt{m+1}(|i,m+1\rangle\langle i,m| + |i,m\rangle\langle i,m+1|) \qquad (10)$$

which for each molecular energy level $i$ is analogous to tight-binding model with different site energies and site-to-site hopping integrals (see Fig.1). Finally, the tunneling part can also be rewritten in terms of considered basis set as:

$$\tilde{H}_T = \sum_{k\in\alpha;i,m}(\gamma_{k,i}^m|k,m\rangle\langle i,m| + h.c.), \qquad (11)$$

where $\gamma_{k,i}^m$ is the coupling between the $m$th channel in the electrode and the molecular system, respectively.

2.2 Determination of thermoelectric characteristics.

To avoid unnecessary complexities, in further analysis we take into account molecular bridge which is represented by one electronic level – generalization to multilevel system is simple. When phonon quanta are present on the dot, an electron entering from the left hand side can suffer inelastic collisions by absorbing or emitting phonons before entering the right electrode. Such processes are presented graphically in Fig.1, where individual channels are indexed by the number of phonon quanta in the left $m$ and right electrode $n$, respectively. Each of the possible processes is described by its own transmission probability, which can be written in the factorized form:

$$T_{m,n}(\varepsilon) = \Gamma_L \Gamma_R \left| G_{m+1,n+1}(\varepsilon) \right|^2. \tag{12}$$

Such transmission function (Eq.(12)) is expressed in terms of the so-called linewidth functions $\Gamma_\alpha$ ($\alpha = L, R$) and the matrix element of the Green's function defined as:

$$G(\varepsilon) = \left[ 1\varepsilon - \tilde{H}_M - \Sigma_L - \Sigma_R \right]^{-1}. \tag{13}$$

Here: 1 stands for identity matrix, $\tilde{H}_M$ is the molecular Hamiltonian (given by Eq.(10)), while the effect of the electronic coupling to the electrodes is fully described by specifying self-energy corrections $\Sigma_\alpha$.

In the present paper we adopt the wide-band approximation to treat metallic electrodes, where the hopping matrix element is independent of energy and bias voltage, i.e. $\gamma_{k,i}^m = \gamma_\alpha$. In this case, the self-energy is given through the relation:

$$\Sigma_\alpha = -\frac{i}{2}\Gamma_\alpha, \tag{14}$$

where

$$\Gamma_\alpha = 2\pi |\gamma_\alpha|^2 \rho_\alpha. \tag{15}$$

Here: $\rho_\alpha$ is the density of states in the $\alpha$-electrode. Both electrodes are also identified with their electrochemical potentials [24]:

$$\mu_L = \varepsilon_F + (1-\eta)eV, \tag{16}$$

and

$$\mu_R = \varepsilon_F - \eta eV \tag{17}$$

which are related to the Fermi energy level $\varepsilon_F$. The voltage division factor $0 \leq \eta \leq 1$ describes how the electrostatic potential difference $V$ is divided between two contacts and can be related to the relative strength of the connection to two electrodes: $\eta = \gamma_L/(\gamma_L + \gamma_R)$. Here we assume the case of asymmetric connection ($\gamma_L < \gamma_R \Rightarrow \eta < 0.5$) in which the rectification effect is generated [24].

In the nonlinear response regime, both the electric current flowing through the junction and the energy flux can be expressed in terms of transmission probability of the individual transitions $T_{m,n}$ which connects incoming channel $m$ with outgoing channel $n$:

$$\begin{pmatrix} I \\ Q \end{pmatrix} = \frac{2}{h} \int_{-\infty}^{+\infty} d\varepsilon \begin{pmatrix} e \\ \varepsilon \end{pmatrix} \sum_{m,n} T_{m,n} \left[ P_m f_L^m (1 - f_R^n) - P_n f_R^n (1 - f_L^m) \right], \tag{18}$$

where

$$f_\alpha^m = [\exp[\beta_\alpha(\varepsilon + m\Omega - \mu_\alpha)] + 1]^{-1} \quad (19)$$

is the Fermi distribution function. Similar expressions for the energy flux can be found in the literature [25-27]. The factor of 2 in Eq.(18) accounts for the two spin orientations of conduction electrons. It is obvious that the elastic contributions to the currents can be obtained from Eq.(18) by imposing the constraint of elastic transitions, where $\varepsilon_{in} = \varepsilon_{out}$ or more precisely $m = n$. It should be noted that the nonlinearity of Eq.(18) is associated with the exponential dependence of the Fermi functions on bias voltage and the exponential dependence of the Fermi and Boltzmann functions on temperatures.

Within the linear response approximation ($V \to 0$, $\delta\theta \to 0$), the charge and energy fluxes are directly proportional to the chemical potential difference $V = \mu_L/e - \mu_R/e$ and the temperature difference $\delta\theta = \theta_L - \theta_R$ and therefore we can write down [27-29]:

$$\begin{pmatrix} I \\ Q \end{pmatrix} = L \begin{pmatrix} V \\ \delta\theta \end{pmatrix}. \quad (20)$$

Here thermoelectric coefficients are given through the following expressions:

$$L = \begin{bmatrix} e^2 J_0 & ek_B\beta_R J_1 \\ eJ_1 & k_B\beta_R J_2 \end{bmatrix}. \quad (21)$$

Obviously, the so-called Onsager relation is fulfilled: $L_{21} = L_{12}/k_B\beta_R$, while the integrals $J$ are defined as the following convolutions:

$$J_x = \frac{2}{h} \int_{-\infty}^{+\infty} d\varepsilon' \sum_{m,n} [-\varepsilon']^x F_T(\varepsilon - \varepsilon') P_m T_{m,n}(\varepsilon'), \quad (22)$$

for $x = 0, 1, 2$ and with the so-called thermal broadening function:

$$F_T(\varepsilon) \equiv -\frac{\partial f}{\partial \varepsilon} = \frac{\beta_R}{4} \operatorname{sech}^2\left(\frac{\beta_R}{2}\varepsilon\right). \quad (23)$$

Such function is responsible for the broadening of the peaks of thermoelectric coefficients. Since typical thermal energy scale ($\beta_R^{-1} \leq 0.03 eV$) is relatively small in comparison with molecular transport scale ($\sim eV$), we can usually approximate thermal broadening function with the help of Dirac delta function: $F_T(\varepsilon) \cong \delta_D(\varepsilon)$.

## 3. Results and discussion

To illustrate that method in a simplest possible context, in further analysis we consider one electronic level $\varepsilon_0$ which is connected to two metallic electrodes (described within wide-band approximation), where the electrons on the dot are coupled with the coupling strength $\lambda$ to a single phonon mode with energy $\Omega$ (primary mode). Generalization to multilevel system with many different phonons can be obtained straightforwardly. Here we adopt the following parameters of the model (given in eV): $\varepsilon_0 = 0$ (the reference LUMO energy), $\varepsilon_F = -1$, $\Omega = 1$, $\lambda = 0.5$, $\rho_L^{-1} = \rho_R^{-1} = 14.3$ ($\rho = 0.07/eV$ is suitable for gold electrodes), $3\gamma_L = \gamma_R = 0.24$ (asymmetric connection), $\beta_L^{-1} = 0.015$, $\beta_R^{-1} = 0.025$ (the room temperature). Asymmetry is maybe not too strong, but for the case of $\gamma_L \ll \gamma_R$ we get: $\eta \to 0$, voltage drop takes place only at the left contact ($\mu_L = \varepsilon_F + eV$, $\mu_R = \varepsilon_F$), while all the transport characteristics are swept out for negative biases. Moreover, we choose the maximum number of the allowed phonon quanta $m_{max} = 4$ to obtain fully converged results for all the parameters involved in this paper.

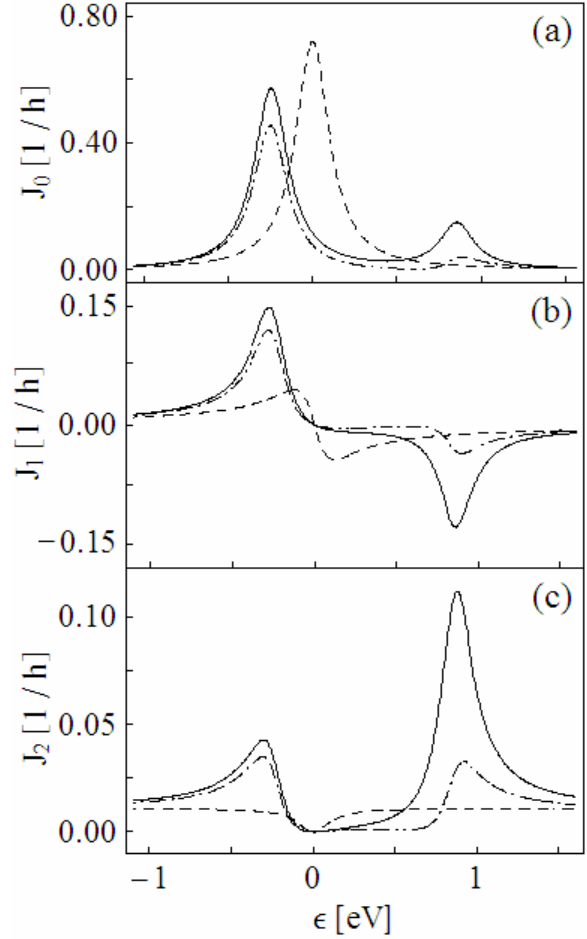

Figure 2: Integrals $J$ associated with particular thermoelectric coefficients as functions of electron energy $\varepsilon$. Total value of the integral (solid line) and its elastic part (dashed-dotted line) are compared with integral calculated in the absence of phonons (dashed line). Parameters of the model are given in the text.

In Fig.2 we plot energy-dependent integrals (given through Eq.(22)) which are related to thermoelectric coefficients. Resonant structure of integral $J_0$ represents the electrical conductance $L_{11}$ (given in units $e^2/h$). When there are no phonons on the molecular bridge, one conductance peak is located symmetrically with respect to the molecular energy level ($\varepsilon_0 = 0$) reaching the value of one-level spin-degenerate transmission:

$$\max(J_0) = \frac{8\Gamma_L \Gamma_R}{(\Gamma_L + \Gamma_R)^2} = 0.72 . \qquad (24)$$

In the presence of electron-phonon coupling, the overall conductance spectrum is shifted by $\Delta = -\lambda^2/\Omega = -0.25$ due to the polaron formation. In addition, the main resonant peak is reduced in height (below $e^2/h$) and new satellite peaks separated by the phonon energy $\Omega$ appear at positive energy side. The intensity of the satellite peaks is much smaller than the main resonant peak because they are evolving from the emission of phonons, which is controlled with the help of the $\lambda$-parameter. Positions of the peaks approximately coincide with polaron energies:

$$\varepsilon_{pol}(m) = \varepsilon_0 - \frac{\lambda^2}{\Omega} + m\Omega, \qquad (25)$$

where $m$ denotes the $m$ th excited state of a polaron (defined as a state of an electron coupled to phonons). The mentioned conductance peaks are of Lorentzian shape and their height depends on three factors: the average temperature of the system, the phonon energy $\Omega$, and the strength of electron-phonon interaction $\lambda$. Integral $J_1$ is associated with coefficients $L_{12}$ and $L_{21}$ (given in $ek_B\beta_R/h$ and $e/h$, respectively). Such quantity changes its sign when passing through the molecular energy level ($\varepsilon_0 = 0$). Besides, in the presence of phonons, the values of $J_1$ are significantly enhanced in the vicinity of polaron resonances. Positively-defined integral $J_2$ contains an information concerning the coefficient $L_{22}$ (given in $k_B\beta_R/h$). In non-phonon case, discussed quantity is everywhere relatively small with minimum exactly at the resonance ($\varepsilon_0 = 0$). Inclusion of electron-phonon interaction leads to occurrence of distinct peaks coincided with polaron resonances, but this time the intensity of the first satellite peak is much bigger than the main resonant peak. Interestingly, all the peaks of the thermoelectric coefficients have as well elastic as inelastic contribution.

The dependence of the electrical current and nonlinear conductance on bias voltage is demonstrated in Figs.3a and 3b, respectively. The maximal current flowing through the junction (given in units $e/h$) can be estimated from the relation:

$$\max(I) = 4\pi \frac{\Gamma_L \Gamma_R}{\Gamma_L + \Gamma_R} \cong 0.29. \qquad (26)$$

Molecular vibrations are observed with different intensities in the positive and negative bias polarity, thus for clarity we focus our attention on the positive bias region in the spectrum. The calculations successfully reproduce typical features of the I-V characteristics commonly observed in inelastic electron tunnelling spectroscopy (IETS experiments) [18-23]. Here we recognize polaron shift of current steps in the direction to lower voltages, which is responsible for a reduction of the conductance gap. It should be mentioned that this gap can be also increased due to polaron shift and the final result depends on the effect of polaronic modification of the thermoelectric coefficient $L_{11}$ in relation to the Fermi level. Obviously, an additional current step is associated with the first excited polaron state.

In Fig.3c the calculated energy flux is displayed against the applied bias. Here we can also find the traces of polaronic transport. Besides, an amount of energy transferred in the presence of electron-phonon interactions is usually much bigger than the one calculated for non-phonon case. The sign of the energy flux depends mainly on the conduction nature. If transport is due to the electron conduction through the LOMO level (as in our case), the heat current achieves negative values for positive bias voltage in the absence of phonons. From the other hand, if transport is due to the hole conduction through the HOMO level, the heat current achieves positive values for positive bias voltage. However, we can not formulate such general conclusions for the situation, where transport is associated with virtual polaron states – analysis in this case is much more complicated.

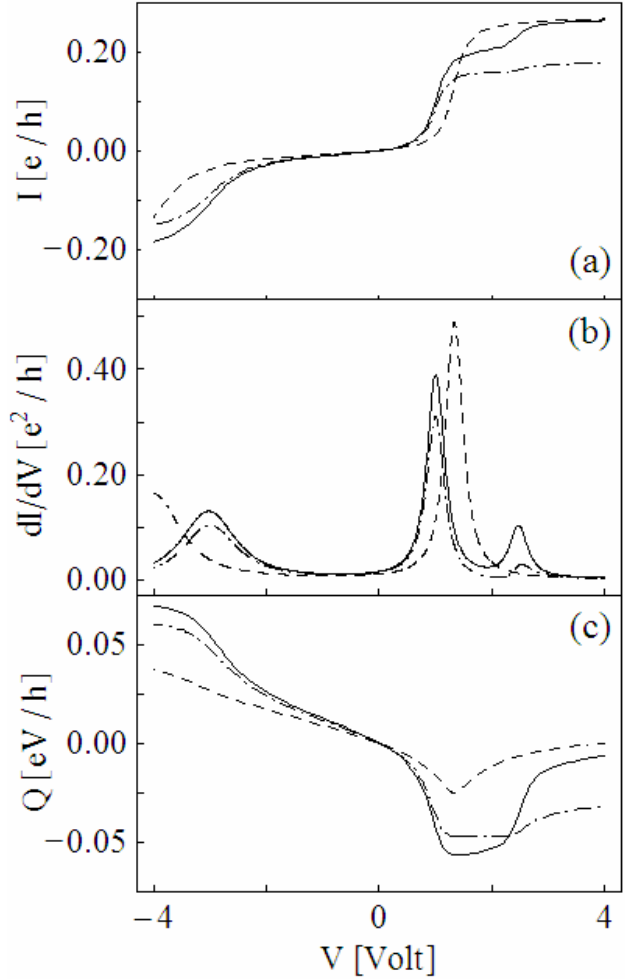

Figure 3: The electrical current (a), its first derivative with respect to voltage and the energy flux (c) plotted against the bias voltage. Total flux (solid line) and its elastic part (dashed-dotted line) are compared with the flux obtained in the absence of phonons (dashed line). Parameters of the model are given in the text.

## 4. Concluding remarks

Summarizing, we have presented a general method that can be successfully used to study the thermoelectric effects of polaronic transport through molecular junctions, composed of vibrating molecular bridges connected to the metallic electrodes. This nonperturbative computational scheme is based on the Green's function theory within the framework of polaron transformation which maps the many-body electron-phonon interaction problem exactly into a one-body multi-channel single-electron scattering problem. It is appropriate to remark that an approximate mapping to a one-electron picture is well justified in a high-bias limit. In the present method, we have completely ignored the following effects: phase decoherence processes in the treatment of the electron-phonon exchange, Coulomb interactions between charge carriers, and the phonon mediated electron-electron interactions. However, an obvious advantage of this technique is associated with the fact that it does not involve any restrictions on the parameters of the model.

It is worth to mention that having all the thermoelectric coefficients we can also find other important characteristics. For the case of zero electric transport current ($I = 0$), the thermopower $S$ can be found by measuring the induced voltage drop across the junction when a small temperature difference between two reservoirs is applied ($V = S \times \delta\theta$). Thus the thermopower can be defined through the following relation:

$$S \equiv \lim_{\delta\theta \to 0} \frac{V}{\delta\theta}\bigg|_{I=0} = -\frac{L_{12}}{L_{11}} = -\frac{k_B \beta_R J_1}{e J_0} \qquad (27)$$

with $k_B/e \cong 86.17 \mu V/K$. Under the mentioned condition ($I = 0$), the heat flux is directly proportional to temperature difference: $Q = \kappa \times \delta\theta$, where the thermal conductance $\kappa = -S L_{21} + L_{22}$. Besides, it is also useful to define dimensionless quantity $\Delta \equiv L_{12} L_{21} / L_{11} L_{22}$ which provides a measure of the rate of entropy production in the transport process ($0 \leq \Delta \leq 1$) [30]. However, the detailed discussion of such quantities is beyond the scope of this work.

Here we have shown that polaronic effects have an essential influence on considered transport characteristics in the linear and nonlinear response regimes. Anyway, inelastic transport is quite important for the structural stability and the switching possibility of the molecular electronic devices. Recently, the polaron formation on the molecule was also suggested as a possible mechanism for generating the hysteretic behaviour of the $I-V$ dependence [31-33] and negative differential resistance (NDR effect) [33]. Inelastic electron tunnelling spectroscopy (IETS) not only helps in understanding of the vibronic coupling between the charge carriers and the nuclear motion of the molecule, but also provides a powerful tool for studying the geometrical structures in molecular junctions. In particular, it has been shown that the spectra of molecular junctions with different geometries have very different spectral profiles [22,23].

**Acknowledgement**

The author is very grateful to the Referee of this article for careful discussions and valuable suggestions.**References**

[1] C. Joachim, J.K. Gimzewski and A. Aviram, Nature (London) **408**, 541 (2000).
[2] A. Nitzan and M.A. Ratner, Science **300**, 1384 (2003).
[3] J.R. Heath and M.A. Ratner, Phys. Today **56**, 43 (2003).
[4] C. Joachim and M.A. Ratner, Proc. Natl. Acad. Sci. USA **102**, 8801 (2005).
[5] K. Haule and J. Bonča, Phys. Rev. B **59**, 13087 (1999).
[6] E.G. Emberly and G. Kirczenow, Phys. Rev. B **61**, 5740 (2000).
[7] H. Ness and A.J. Fisher, Chem. Phys. **281**, 279 (2002).
[8] M. Čížek, M. Thoss and W. Domcke, Phys. Rev. B **70**, 125406 (2004).
[9] K. Walczak, Physica E **33**, 110 (2006).
[10] J.-X. Zhu and A.V. Balatsky, Phys. Rev. B **67**, 165326 (2003).
[11] B. Dong, H.L. Cui, X.L. Lei and N.J.M. Horing, Phys. Rev. B **71**, 045331 (2005).
[12] K. Walczak, J. Magn. Magn. Mater. **305**, 475 (2006).
[13] M. Paulsson and S. Datta, Phys. Rev. B **67**, 241403 (2003).
[14] J. Koch, F. von Oppen, Y. Oreg and E. Sela, Phys. Rev. B **70**, 195107 (2004).10